\documentclass[%
reprint,
 amsmath,amssymb,
 aps,
prb,
]{revtex4-2}

\usepackage{graphicx}
\usepackage{dcolumn}
\usepackage{bm}
\usepackage{hyperref}
\usepackage{float}
\usepackage[dvipsnames]{xcolor}

\begin{document}
\newcommand{\revsapta}[1]{{\color{red}{#1}}}
\preprint{APS/123-QED}

\title{Modeling Interlayer Interactions and Phonon Thermal Transport in Silicene Bilayer}

\author{Sapta Sindhu Paul Chowdhury}
\author{Appalakondaiah Samudrala}
\author{Santosh Mogurampelly}%
\email{santosh@iitj.ac.in}
\affiliation{%
 Department of Physics, Indian Institute of Technology Jodhpur\\
 Karwar, Jodhpur, Rajasthan, India - 342030. 
}%




\date{\today}

\begin{abstract}
We develop an accurate interlayer pairwise potential derived from the \textit{ab-initio} calculations and investigate the thermal transport of silicene bilayers within the framework of equilibrium molecular dynamics simulations. The electronic properties are found to be sensitive to the temperature with the opening of the band gap in the $\Gamma$$\rightarrow$M direction. The calculated phonon thermal conductivity of bilayer silicene is surprisingly higher than that of monolayer silicene, contrary to the trends reported for other classes of 2D materials like graphene and hBN bilayers. This counterintuitive behavior of the bilayer silicene is attributed to the interlayer interaction effects and inherent buckling, which lead to a higher group velocity in the LA$_1$/LA$_2$ phonon modes. The thermal conductivity of both the mono- and bilayer silicene decreases with temperature as $\kappa\sim T^{-0.9}$ because of the strong correlations between the characteristic timescales of heat current autocorrelation function and temperature ($\tau\sim T^{-0.75}$). The mechanisms underlying phonon thermal transport in silicene bilayers are further established by analyzing the temperature induced changes in acoustic group velocity.

\end{abstract}

\maketitle


\section{\label{sec:introduction}Introduction\protect}

Silicene is an intriguing two dimensional (2D) material with unique in-plane atomic configuration analogous to graphene and out-of-plane structural similarities with other classes of 2D materials \cite{Vogt2012, Lin2013, Kharadi2020}. The structural uniqueness inherent to silicene offers several distinguishable physical and electronic properties \cite{Fleurence2012}. Silicene has garnered tremendous interest in the last decade due to the existence of remarkable flexural phonon scattering \cite{Tao2015}, electro-optic effects \cite{Matthes2013}, quantum anomalous Hall effect \cite{Wu2014}, spin-orbit couplings \cite{Liu2011}, giant magnetoresistance \cite{Xu2012}, and tunable band structure \cite{Drummond2012}. The flexural phonon scattering in silicene limits the thermal conductivity to low values, opening new avenues for technological applications such as thermoelectric devices. More interestingly, the structural and compositional superiority makes silicene compatible with silicene based electronics \cite{Shan2023} and state-of-the-art semiconductor fabrication technologies \cite{Novoselov2019}.

The bilayer silicene poses interesting challenges for technological applications \cite{Pflugradt2014, Sakai2015, Ouyang2018, Shih2019}, and therefore, understanding the underlying physical mechanisms is essential. A majority of the investigations carried out so far focused on the electronic properties of bilayer silicene \cite{Liu2013, Zhang_2015, Do2018, Do2019}. Despite significant progress in elucidating the electronic properties of mono- and bilayer silicene 2D materials, the phonon thermal transport and the underlying fundamental heat transport mechanisms are poorly understood \cite{spencer_morishita_2018}. Specifically, while there exist only a few computational works reporting the thermal conductivity of monolayer silicene, there is little work \cite{ZHOU2018} on the thermal conductivity of the bilayer. 


Understanding the thermal transport mechanisms in unique 2D materials like silicene has fundamental importance from both the scientific and the technological points-of-view \cite{Gu2018}. Technological application of these 2D materials in thermal interfacing applications, thermoelectric devices require thorough understanding of phonon transport at higher temperatures \cite{Xiong2010pre}. As a result, both theoretical and experimental efforts have been undertaken to understand the thermal transport in these materials. Thermal conductivity of monolayer, bilayer and twisted bilayers of graphene, h-BN etc. have been studied extensively with monolayer graphene showing the highest thermal conductivity. Similarly, many works have focused on the dependency of thermal conductivity with the external factors like strain, temperature, twist etc. The thermal conductivity in silicene and the interplay of interactions is not well understood due to the challenges involved in modeling these materials. \cite{Wang2012pre, Xiong2010pre}.

The computational studies of monolayer silicene focused on the thermal conductivity and its dependency on temperature \cite{Pei2014}, strain \cite{Xie2016, Kuang2016}, vacancy \cite{Li2012}, and substrate \cite{Wang2015}. The first principles based calculations estimated the thermal conductivity of silicene to be 9.4 W/(m K) at room temperature, mainly due to the in-plane vibrations \cite{Xie2014}. Using the reactive force field (ReaxFF) model, molecular dynamics (MD) study found the silicene monolayer to be thermally stable up to a temperature of 1500 K \cite{Berdiyorov2014}. Earlier work based on the Tersoff bond order potential \cite{Tersoff1989} based MD simulations estimated the thermal conductivity to be 20 W/(m K) \cite{Hu2013} but failed to reproduce the buckled structure of silicene. Similarly, the Modified Embedded Atom Model (MEAM) potential incorrectly predicts the buckling height (twice the reported value) in silicene \cite{Pei2013}. Stillinger-Weber (SW) potential parameters modified by Zhang et al. \cite{swopt2014} could reproduce the buckling of silicene monolayer in agreement with experiments and first principle density functional theory (DFT) studies. The modified SW potential parameter set produces the thermal conductivity of the monolayer to be in the range 5-15 W/(m K) at 300 K, employing both equilibrium and non-equilibrium molecular dynamics (EMD and NEMD) approaches \cite{swopt2014}.

Despite the above, there is no consensus on how the thermal conductivity of monolayer silicene scales with temperature and the corresponding phonon mechanisms. More importantly, the investigations on phonon thermal properties of multilayered silicene suffer a serious lack of the availability of a suitable interlayer potential. In this work, we optimize the Lennard-Jones (LJ) pairwise interaction potential to accurately capture the interlayer interactions using the \textit{ab-initio} DFT calculations. The optimized interlayer LJ potential model is used for studying the structure and thermal transport properties of the silicene bilayer. Specifically, we study the effect of temperature on the structural, electronic, and thermal conducting properties of the silicene bilayer. Moreover, we compare the thermal transport of the silicene bilayer to that of the monolayer. The thermal conductivity results of silicene are explained through the calculation of mode dependent phonon properties of the bilayer and monolayer systems. The phonon characteristic timescales associated with the heat current autocorrelation functions and the phonon dynamics are thoroughly analyzed to examine the origins behind the power-law scaling relation between thermal conductivity and temperature.

The rest of the paper is organized as follows: Section \ref{sec:theory} presents the theoretical formalism and computational details followed in the paper. Section \ref{sec:result} presents the results obtained in the study, followed by a discussion, and Section \ref{sec:conclusions} summarizes the key results of the paper. We present the structural information in Section \ref{sec:structure}, and the optimization of LJ parameters is detailed in Section \ref{sec:optimization}. The effect of temperature on the structural configurations is described in Section \ref{subsec:temp_effect} followed by the calculations of thermal conductivity in Section \ref{subsec:cond}. Section \ref{subsec:phonon} explains the thermal conductivity by analyzing phonon modes.

\section{\label{sec:theory}Theoretical Background and Computational Details\protect}

We use the \emph{ab-initio} density functional theory framework as implemented in the 
\textit{quantum espresso} \cite{Giannozzi2007,Giannozzi2017} to predict the ground-state interlayer separation between silicene layers. The exchange–correlation (XC) energy was estimated using the Perdew–Burke–Ernzerhof (PBE) functional \cite{pbe1996} of the generalized gradient approximation (GGA). It is noteworthy mentioning that GGA-PBE functional commonly leads to an underestimation of the van der Waals (vdW) interactions in layered materials like graphene and h-BN, where the $\pi$-bonding network plays a critical in determining their properties. However, the interactions in bilayer or multi-layered silicene materials are predominantly governed by strong covalent bonds that disrupt the $\pi$-bonding network between the layers, and therefore, integrating the empirical vdW correction schemes into our DFT calculations is unnecessary (more details are available in the Supplemental Material \cite{supplement}. See also references \cite{Wang2017,Fu2014,Padilha2015} therein). The projector augmented wave (PAW) \cite{paw1999} method for pseudopotential is used to address the electron core interactions. For our calculations, plane waves with a kinetic energy cutoff of 110 Ry for wavefunction and 880 Ry for charge density have been used with a Monkhorst-Pack \cite{mp1976} k-point grid of 32 $\times$ 32 $\times$ 1. We have minimized the Hellmann-Feynman forces acting on the atoms to $5\times10^{-4}$ eV/{\AA}, and the total energy convergence threshold is $10^{-6}$ eV.

The molecular dynamics simulations have been carried out in the Large-scale Atomic/Molecular Massively Parallel Simulator (Lammps) \cite{LAMMPS}. In our approach, we have employed SW \cite{sw1985} potential  to describe the \textit{intralayer} interactions, while the \textit{interlayer} interactions are modeled using the LJ potential. Specifically, the atoms within a given layer interact exclusively through the SW potential, and the atoms of different layers interact solely via the LJ potential. The parameters governing the SW potential have been adopted from Zhang et al. \cite{swopt2014}, while the parameters for the LJ model are subject to refinement within the scope of this study. The Hamiltonian used in our classical physics calculations is given by:
\begin{equation}
	H = \text{K.E.} + \Phi_{\text {intra}} + \Phi_{\text {inter}},
\end{equation} 
where the terms on the right-hand side represent the kinetic energy, \textit{intralayer} interaction potential, and \textit{interlayer} interaction potential of the system, respectively. The term, $\Phi_{\text {intra }}$ takes the following form:
\begin{eqnarray}	
\Phi_{\text {intra }}=&&\sum_i \sum_{j>i} \sum_l \phi_2\left(r_{ijl}\right) \nonumber\\
&&+\sum_i \sum_{j \neq i} \sum_{k>j} \sum_l \phi_3\left(r_{ijl}, r_{ikl}, \theta_{ijkl}\right).
\end{eqnarray} 
In the above equation, $r_{ijl}$ represents the distance between $i^{\text{th}}$ and $j^{\text {th}}$ atom of $l^{\text{th}}$ layer and $\theta_{ijkl}$ represents the angle between $i^{\text{th}}$, $j^{\text{th}}$, and $k^{\text{th}}$ atom of $l^{\text{th}}$ layer. The terms, $\phi_2$ and $\phi_3$ are respectively given by:
\begin{eqnarray}
\phi_2\left(r_{ijl}\right)=&&A_{ijl} \epsilon_{i j l}\left[B_{i j l}\left(\frac{\sigma_{i j l}}{r_{i j l}}\right)^{p_{i j l}}-\left(\frac{\sigma_{i j l}}{r_{i j l}}\right)^{q_{i j l}}\right] \nonumber\\
&&\times\exp \left(\frac{\sigma_{i j l}}{r_{i j l}-a_{i j l} \sigma_{i j l}}\right)
\end{eqnarray} 
and
\begin{eqnarray}
\phi_3\left(r_{i j l}, r_{i k l}, \theta_{i j k l}\right)=&&\lambda_{i j k l} \epsilon_{i j k l}\left[\cos \theta_{i j k l}-\cos \theta_{0 i j k l}\right]^2 \nonumber\\
&&\times\exp \left(\frac{\gamma_{i j l} \sigma_{i j l}}{r_{i j l}-a_{i j l} \sigma_{i j l}}\right) \nonumber\\
&&\times\exp \left(\frac{\gamma_{i k l} \sigma_{i k l}}{r_{i k l}-a_{i k l} \sigma_{i k l}}\right).
\end{eqnarray} 
Here, $a, \lambda, \gamma, \cos \theta_0, A, B, p, q$ are the force field parameters defining the interaction potential terms, taken from Zhang et al. \cite{swopt2014}. The \textit{interlayer} potential, $\Phi_{\text {inter}}$ is of the Lennard-Jones form:
\begin{equation}
\Phi_{\text {inter}}=\sum_{i \in L_1} \sum_{j \in L_2} \phi\left(r_{i j}\right),
\end{equation} 
where,
\begin{equation}
\phi\left(r_{i j}\right)=4 \epsilon\left[\left(\frac{\sigma}{r_{i j}}\right)^{12}-\left(\frac{\sigma}{r_{i j}}\right)^6\right], r_{i j}<r_c.
\end{equation}
Here, $\sigma$ and $\epsilon$ are the parameters of the potential, $r_{ij}$ is the distance between $i^{\text{th}}$ and $j^{\text{th}}$ atoms of different layers. The global cutoff for LJ is chosen to be $r_c = 20 $ {\AA}, which is more than five times $\sigma$. 
For optimization of the LJ parameters for bilayer silicene, we have calculated the binding energy using the equations: $E^{\mathrm{DFT/MD}}_\mathrm{b}$ = $E^\mathrm{{DFT/MD}}_\mathrm{{bilayer}}$ - $E^\mathrm{{DFT/MD}}_\mathrm{{layer1}} - E^\mathrm{{DFT/MD}}_\mathrm{{layer2}}$ where, $E^\mathrm{{DFT}}_\mathrm{{b}}$ and $E^\mathrm{{MD}}_\mathrm{{b}}$ are the binding energy from DFT and MD calculations respectively. Here, $E^\mathrm{{DFT}}_\mathrm{bilayer}$, $E^\mathrm{{DFT}}_\mathrm{layer1}$ and $E^\mathrm{{DFT}}_\mathrm{layer2}$ are the total energies of the bilayer, and the monolayers from DFT calculations and $E^\mathrm{{MD}}_\mathrm{bilayer}$, $E^\mathrm{{MD}}_\mathrm{layer1}$ and $E^\mathrm{{MD}}_\mathrm{layer2}$ are the total energies of the bilayer and the monolayer from MD simulations. We minimize the following objective function to get the optimized $\sigma$ and $\epsilon$:
\begin{equation}
	\chi^2 = \frac{1}{W}\int_{0}^{\infty}|E^{\mathrm{DFT}}_\mathrm{b}(r) - E^{\mathrm{MD}}_\mathrm{b}(r)|^2 w(r)dr
\end{equation}
\begin{equation}
	\label{eqn:chisq}
	 \approx \frac{1}{W}\sum_{r_i}^{r_{cut}}|E^{\mathrm{DFT}}_\mathrm{b}(r_i) - E^{\mathrm{MD}}_\mathrm{b} (r_i)|^2 w(r_i),
\end{equation}
where, $r$ specifies the interlayer separation between two silicene layers and $w(r_i)$ is the weight function, defined as:
\begin{equation}
	w(r) = \exp\left[-\left(\frac{r - r_0}{\zeta}\right)^2\right],
\end{equation}
where, $r_0$ is the minima of LJ potential and $\zeta$ controls the amplitude of $w(r)$ around the minima. We note that the choice of $w(r)$ is not unique and a relevant weight function can be chosen so that the significant part of the LJ potential is represented correctly.

We have applied the periodic boundary condition (PBC) in all three directions. To create aperiodicity and to prevent the interactions between the bilayer system and its images in the z-direction, we have created 50 {\AA} of vacuum in the simulation cell. Energy minimization using the conjugate gradient (CG) \cite{cg1952} and steepest descent (SG) algorithms have been conducted with energy tolerance of $10^{-8}$ and force tolerance of $10^{-8}$ eV/{\AA}.
The minimized structures have been equilibrated in the isothermal-isobaric (NPT) ensemble at different temperatures starting from 300 K to 1000 K and at 0 bar pressure for 300 ps. The Nos\'e-Hoover thermostat \cite{nose1984, hoover1985} and the Nos\'e-Hoover barostat \cite{nhbaro1983} with a temperature coupling constant of 0.1 ps and pressure coupling constant of 1 ps have been used for maintaining a constant temperature and pressure, respectively. The systems are then equilibrated in a canonical (NVT) ensemble for 300 ps. The Verlet algorithm \cite{verlet1967} with timestep 1 fs has been used to integrate the equation of motion.

For the generation of production trajectories for the calculations of thermal conductivity, we simulated larger trajectories in an NVT ensemble. A total of 30 independent trajectories are generated using uncorrelated initial conditions. Each independent trajectory of length 2 ns is generated using a timestep of 0.1 fs.

We have used the Green-Kubo (GK) relation based on the fluctuation-dissipation theorem to calculate the thermal conductivity of the systems using the following relation:
\begin{equation}
	\label{eqn:gk_cond}
	\bm{\kappa}(T) = \frac{1}{Vk_BT^2}\int_{0}^{\infty} \langle \bm{S}(0).\bm{S}(t) \rangle dt,
\end{equation}
where, $V$ is the volume, $k_B$ is the Boltzmann constant, and $T$ is the temperature of the system. The ensemble average $\langle \bm{S}(0).\bm{S}(t) \rangle$ represents the autocorrelation function of the heat current operator $\bm{S(t)}$, which is calculated from the simulation data as:
\begin{equation}
	\bm{S}(t) = \frac{d}{dt}\sum_{i}\bm{r}_i\tilde{E}_i,
\end{equation}
where $\bm{r}_i$ is the position vector concerning the i\textsuperscript{th} atom, $\tilde{E}_i = E_i - \langle E_i \rangle$ represents the fluctuation in total energy with respect to the mean energy. $\bm{S}(t)$ is numerically computed in the following way:
\begin{equation}
	\bm{S}(t) = \sum_{i}\tilde{E}_i\bm{v}_i + \frac{1}{2}\sum_{i<j}(\bm{F}_{ij}. (\bm{v}_i +\bm{v}_j))\bm{r}_{ij},
\end{equation}
where $\bm{F}_{ij}$ is the force between atom $i$ and atom $j$, $\bm{v}_i$ is the velocity and $\bm{r}_{ij}$ is the displacement of i\textsuperscript{th} and j\textsuperscript{th} atoms. 

The phonon density of states (DoS) of the systems are calculated at different temperatures using the velocity autocorrelation function (VACF) as follows:
\begin{equation}
	\label{eqn:dw}
	D(\omega) = \frac{1}{3Nk_BT}\int_{0}^{\infty}\frac{\langle \bm{v}(0).\bm{v}(t)\rangle}{\langle \bm{v}(0).\bm{v}(0)\rangle}e^{i\omega t}dt,
\end{equation}
where $\langle \bm{v}(0).\bm{v}(t)\rangle$ defines the VACF, $\omega$ is the angular frequency, $N$ is the number of atoms in the system. Notably, a simulation trajectory of 250 ps with a finer timestep of 0.05 fs and a saving frequency of 0.2 fs is used for the NVT run to capture higher resolution in VACF and obtain a smoother DoS. The trajectory is recorded at a time interval of 0.2 fs.

We use the equilibrium statistical mechanics formalism based on the fluctuation-dissipation theorem to construct the dynamical matrices \cite{Kong2011} of phonon wave vector using the following equation:
\begin{equation}
	\bm{D}_{k\alpha,k^{\prime}\beta}(\bm{q}) = (m_km_{k^{\prime}})^{-\frac{1}{2}}\bm{\Phi}_{k\alpha,k^{\prime}\beta}(\bm{q}).
\end{equation}
Here, the force constant coefficient $\bm{\Phi}_{k\alpha,k^{\prime}\beta}(\bm{q})$ at phonon wavevector $\bm{q}$ is obtained from:
\begin{equation}
	\bm{\Phi}_{k\alpha,k^{\prime}\beta}(\bm{q}) = k_{B}T\bm{G}^{-1}_{k\alpha,k^{\prime}\beta}(\bm{q}).
\end{equation}
If the displacement of the k\textsuperscript{th} basis atom in the unit cell in the $\alpha$ direction is:
\begin{equation}
	\bm{u}_{k\alpha} (\bm{q}) = \sum_{l}\bm{u}_{lka}\exp(i\bm{q}.\bm{r}_{l}),
\end{equation}
then the Green's function coefficient can be given as:
\begin{equation}
	\bm{G}_{k\alpha,k^{\prime}\beta}(\bm{q}) = \langle \bm{u}_{k\alpha} (\bm{q}) \cdot \bm{u}^{\ast}_{k^{\prime}\alpha} (\bm{q}) \rangle.
\end{equation}
To calculate the phonon dispersion, we have used a post-processing tool, phana \cite{phana} along with the fix-phonon package \cite{Kong2009} as implemented in lammps. The group velocity is calculated as follows:
\begin{equation}\label{xx}
	\begin{split}
		\bm{v}_g (\bm{q}\nu)& =\nabla_q\omega (\bm{q}\nu)\\
		& =\frac{\partial\omega(\bm{q}\nu)}{\partial\bm{q}}\\
		& =\frac{1}{2\omega (\bm{q}\nu)}\frac{\partial[\omega(\bm{q}\nu)]^2}{\partial\bm{q}}\\
		& =\frac{1}{2\omega (\bm{q}\nu)}\left<\bm{e}(\bm{q}\nu)|\frac{\partial D(\bm{q})}{\partial\bm{q}}|\bm{e}(\bm{q}\nu)\right>.
	\end{split}
\end{equation}
Here, $\bm{e}(\bm{q}\nu)$ is the phonon eigenvector at a frequency $\nu$ at $\bm{q}$. We have used phonopy \cite{phonopy,phonopy3p} interfaced with phono-lammps for calculating the group velocity at different temperatures.
\begin{figure}
	\includegraphics[height=6cm,keepaspectratio]{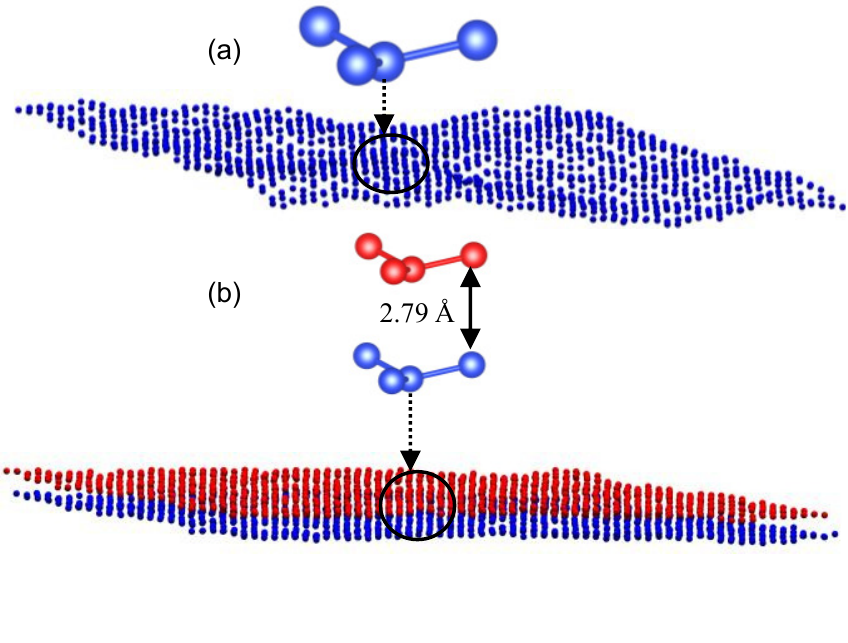}
	\caption{(a) Crystal structure \cite{swopt2014} and the snapshot of equilibrated monolayer silicene displaying out-of-plane corrugations in vacuum at 300 K  (b) crystal structure and the snapshot of equilibrated bilayer silicene displaying out-of-plane corrugations at 300 K. Due to the substrate-like effect on the silicene as a result of interlayer interactions, we observe suppressed spatial fluctuation on the bilayer system compared to the monolayer}
	\label{fig:crystal_structure}
\end{figure}
\section{\label{sec:result}Results and Discussion\protect}
\subsection{\label{sec:structure}Structural Configuration}
\begin{figure}[b]
	\includegraphics[height=6cm, keepaspectratio]{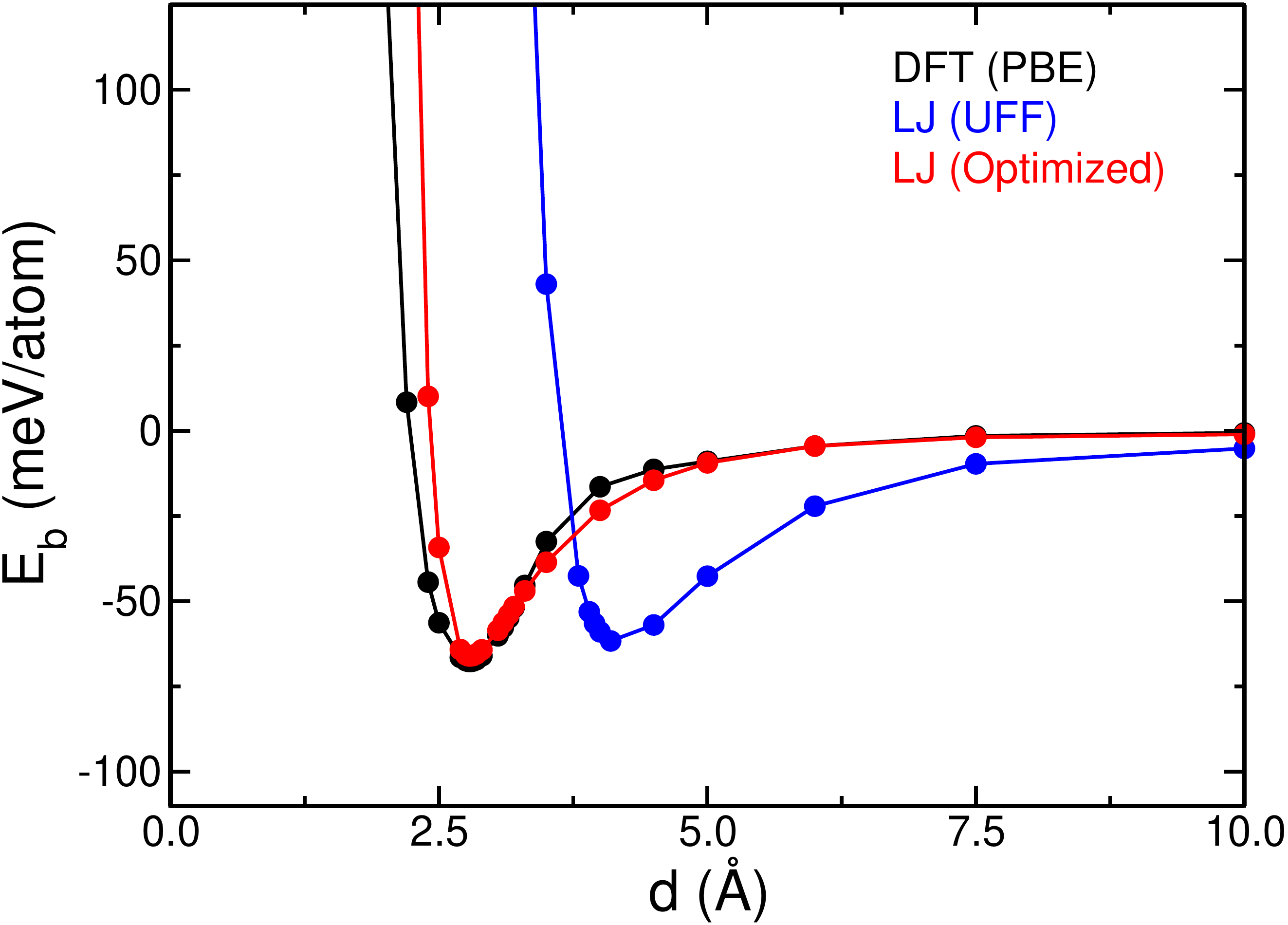}
	\caption{Binding energy with interlayer separation for bilayer silicene. The black curve is the reference DFT data used for deriving LJ parameters. The optimized parameters are used for all subsequent MD simulations. The red-colored curve is the binding energy curve obtained using the optimized LJ parameters ($\epsilon$ = 0.0346 eV and $\sigma$= 2.586 {\AA}).}
	\label{fig:be_curve}
\end{figure}

We consider the low-buckled geometry of monolayer silicene, which is the most stable structure of silicene \cite{spencer_morishita_2018}. The hexagonal lattice with a lattice constant of 3.81 \AA, Si-Si bond length of 2.24 {\AA} with a buckling height of 0.427 {\AA} is taken for our calculation as per previous reports \cite{swopt2014} (Fig. \ref{fig:crystal_structure}(a)). The unit cell follows symmetry operations as followed by the C2/m spacegroup. We extend this unit cell to create a 32 $\times$ 32 $\times$ 1 supercell with 2048 Si atoms (Fig. \ref{fig:crystal_structure}(b)) for subsequent MD simulations. We use the SW2 potential parameters provided by Zhang et al. \cite{swopt2014} to describe interlayer interactions. The parameter set is based on the acoustic part of the phonon dispersion curve responsible for the maximum contribution to thermal conductivity. The optimized geometry obtained using the SD and CG minimization protocols agrees well with the lattice parameters reported. Our DFT calculations confirm the electronic band structure with the Dirac cone at 0 K point and the non-magnetic nature of the monolayer in the ground state \cite{spencer_morishita_2018}. To construct the bilayer, we align one layer on top of the other monolayer so that an AA stacking is obtained. Since no known suitable and accurate interlayer interaction potential exists for bilayer silicene, we develop the LJ potential model. The additional degree of freedom unique to silicene results in a multitude of stacking possibilities, encompassing various buckling and lattice constants. These include AA-planer, AA-buckled, AA$^{\prime}$, AB, AB$^{\prime}$, A$^{\prime}$B$^{\prime}$, which were explored theoretically to comprehend electronic properties \cite{Fu2014, Padilha2015, Pflugradt2014}. Our forthcoming work will focus on developing manybody interlayer interaction potentials tailored for silicene, enabling an understanding of thermal transport in multilayer buckled silicene structures. In the present study, we employ the pairwise potential for the AA-stacked bilayer silicene to study thermal transport, acknowledging the complexity of more comprehensive manybody potentials for diverse stackings.

\subsection{\label{sec:optimization}Optimization of the Force Field Parameters}
\begin{figure}[b]
	\includegraphics[height=6cm,keepaspectratio]{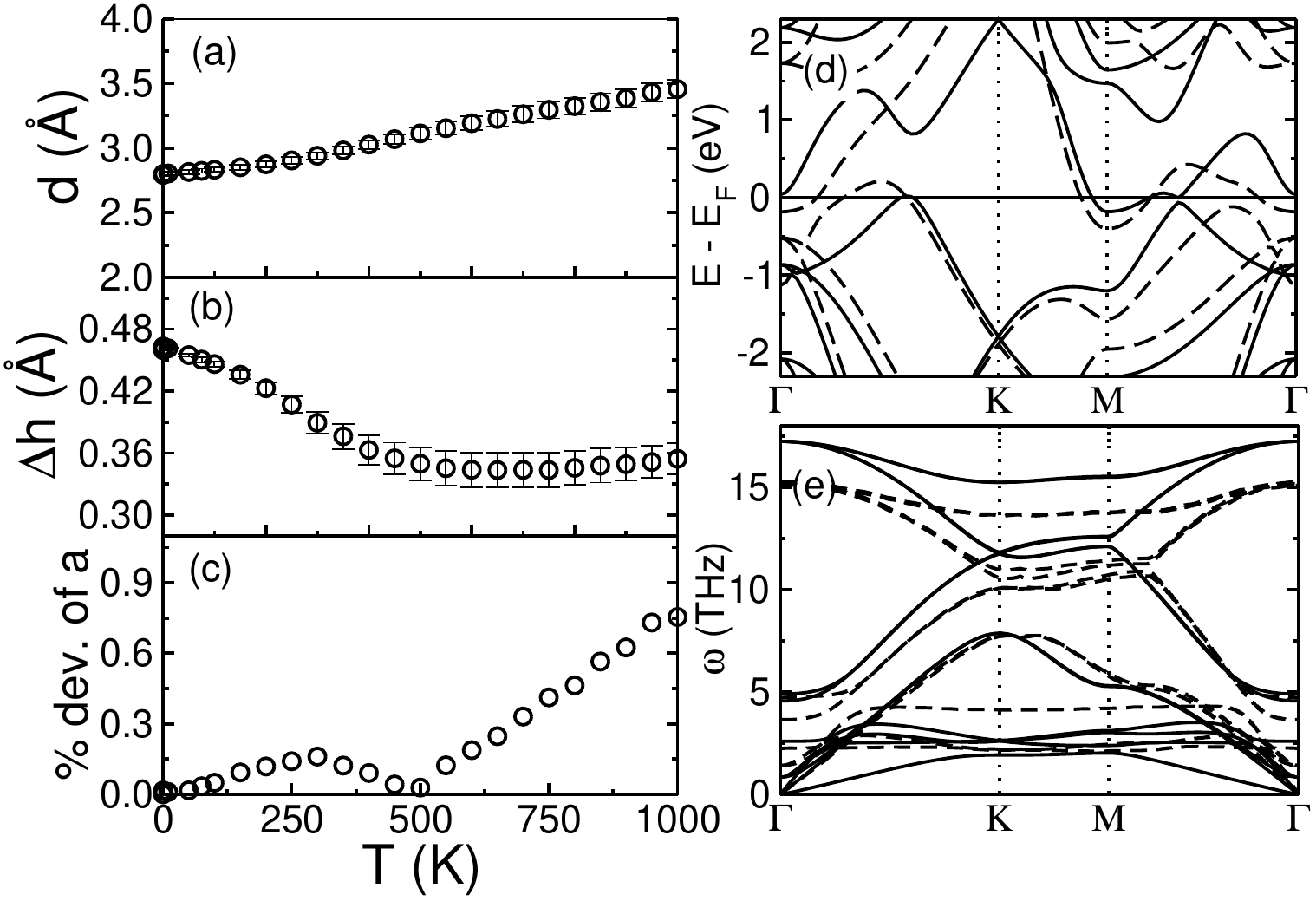}
	\caption{The variation of structural parameters with temperature: (a) interlayer separation, (b) buckling height, and (c) percentage deviation of the lattice constant with respect to the corresponding value at 0 K. (d) The electronic band structure and (e) the phonon dispersion along the  $\Gamma \rightarrow K \rightarrow M \rightarrow \Gamma$ symmetry directions. The solid black line represents the Fermi level in the electronic band. The solid curve represents 0 K, and the dashed curve represents 300 K temperature for both band structures.}
	\label{fig:effect_of_temp}
\end{figure}
\begin{figure}[b]
	\includegraphics[height=12cm,keepaspectratio]{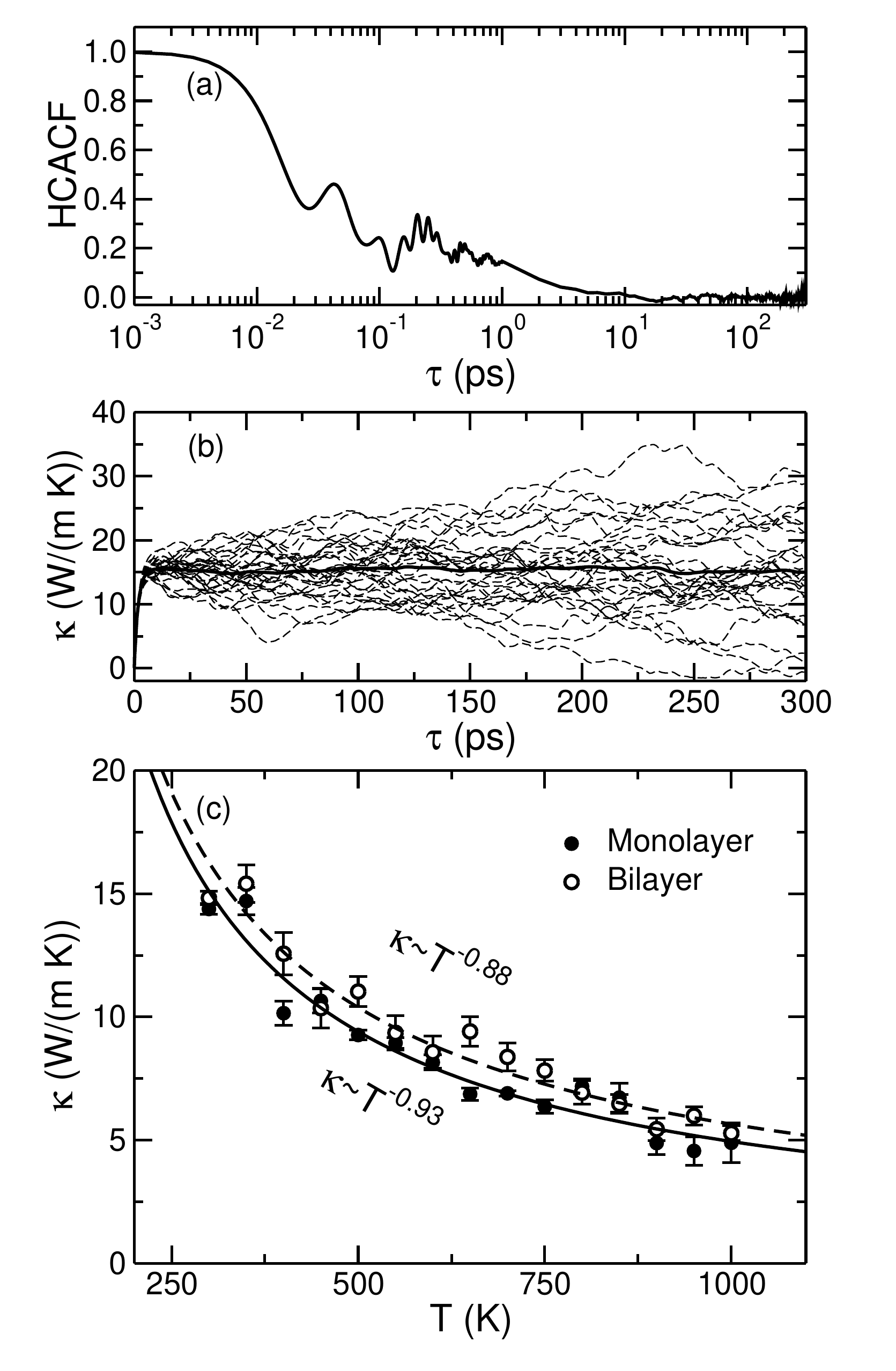}
	\caption{(a) The averaged (over 30 independent runs) normalized HCACF as a function of correlation or lag time at 300 K, (b) the thermal conductivity calculated using 30 independent trajectories (dashed black curve), and the average thermal conductivity (solid black curve) with the upper limit of the Green-Kubo integral (Eq. \ref{eqn:gk_cond}) at 300 K, and (c) the thermal conductivity of silicene mono- and bilayer as a function of temperature. The filled-in circles and the solid line represents the simulation data and the fit to power law $\kappa\sim T^{-0.93}$ for the monolayer system, respectively. The open circles and the dashed line represent the simulation data and the fit to power law $\kappa\sim T^{-0.88}$ for the bilayer system. The power-law fitting shows the decreasing trend of the thermal conductivity with temperature for both the monolayer and bilayer. The bilayer silicene system shows a higher magnitude of thermal conductivity than the monolayer.}
	\label{fig:kappa_vs_T}
\end{figure}
Developing an appropriate potential for modeling the interlayer interactions is a challenging yet crucial task. In the past, many interlayer potentials such as DRIP, KC, and ILP have been developed for graphene and other similar 2D materials \cite{Wen2018,Naik2019,Leven2016}. Though robust and effective in modeling complex geometry such as twisted bilayers and multilayers, simulations with these potentials are computationally highly expensive. Here, we aim to explore the pristine bilayer system without twisting or rotational asymmetry. As such, for simplicity, we develop the pairwise LJ potential to model the interactions between two silicene layers. The reference force field parameters $\epsilon$ and $\sigma$ are based on the unified force field (UFF) model of Rappe et al. \cite{uff1992} and a global cutoff $r_c = 20$ {\AA} which is more than five times $\sigma$ is taken. Previously, in many studies concerning silicene, Si-Si interactions are modeled using the UFF parameters. However, the interlayer separation (d) produced by these parameters is inaccurate. Using the UFF LJ parameters, molecular mechanics minimization produces an interlayer separation of 3.95 {\AA}, which we plan to optimize by tuning $\epsilon$ and $\sigma$. However, our DFT calculations with PBE-GGA density functional show the binding energy is minimum at an interlayer separation of 2.79 {\AA} compatible with previous studies \cite{Padilha2015, Fu2014}. Therefore, we optimize the LJ parameters by taking the UFF as the starting approximation. We have plotted the binding energy curve as a function of the interlayer separation in Fig. \ref{fig:be_curve}. It is clear that the UFF model parameters are far off from our DFT calculation and the literature \cite{Padilha2015, Fu2014}. We obtain the minimized $\epsilon$ and $\sigma$ by optimizing the $\chi^2$ function (Eq. (\ref{eqn:chisq})) following the simplex algorithm developed by Nelder and Mead \cite{nmalgo}. The developed force field parameters for the LJ potential are $\epsilon$ = 0.0346 eV and $\sigma$ = 2.586 {\AA} using weight function parameter $\zeta$ = 0.3 {\AA}. Here, we have introduced the weight function parameter to model the minimum of the binding energy curve accurately. Based on this optimized interlayer potential and the intralayer SW potential, the unit cell of the bilayer considered in this work follows the symmetry operations of the P-3m1 spacegroup. Using the developed LJ force field parameters, the obtained lattice constant is 3.7941 {\AA}, the buckling height is 0.460 {\AA}, and the bond length is 2.238 {\AA}. In Table \ref{tab:table1}, we summarize the structural information of the systems used in our study. Clearly, there is a slight increase in the buckling height while the lattice constant decreases by a bit. We observe that the optimized LJ parameters obtained by us produce the interlayer separation as well as lattice parameters and buckling height of bilayer silicene consistent with the first-principles DFT calculations and in agreement with the literature \cite{Fu2014,Padilha2015} at 0 K. Our DFT calculations with both ferromagnetic and anti-ferromagnetic orderings reveal that the bilayer is most stable in the non-magnetic state. Further details on the validity of our potential is given on Supplementary Material \cite{supplement}(see also references \cite{Naik2019, Silva2020,Wen2018, Liu2013} therein) . 
Despite satisfactorily reproducing the structural features of bilayer silicene correctly, we note that the optimized LJ potential in this work has limited capabilities in the sense that it is most accurate only for the AA-stacked silicene, which it is based on. 
\begin{table}[b]
	\caption{\label{tab:table1}%
		The crystal structures of the monolayer and bilayer silicene in the ground state (0 K) used in our calculations.
	}
	\begin{ruledtabular}
		\begin{tabular}{lcdr}
			\textrm{}&
			\textrm{a ({\AA})}&
			\textrm{ $\Delta h$ ({\AA})}&
			\textrm{Space Group}\\
			\colrule
			Monolayer & 3.812 & 0.427  & C2/m\\
			Bilayer & 3.794 & 0.460 & P-3m1\\
			 
		\end{tabular}
	\end{ruledtabular}
\end{table}

\subsection{\label{subsec:temp_effect}Effect of Temperature on Structural Features}
Because of the buckled topology and structural complexity, silicene is expected to show high sensitivity to temperature. In this section, we discuss the results of structure, electronic and phononic properties as a function of temperature. We use a molecular dynamics trajectory of 500 ps in the isothermal-isobaric ensemble to calculate the structural parameters for the bilayer system at all temperatures. Prior to that, we equilibrate the system in the canonical ensemble for 500 ps at the respective temperatures. The results of temperature dependency of interlayer separation, buckling height, and the change in lattice constant are explored in Fig. \ref{fig:effect_of_temp}. We observe a monotonic increase in the interlayer distance between the layers as the temperature is increased (Fig. \ref{fig:effect_of_temp}(a)). This results from the repulsion between layers due to the increased thermal motion of the atoms at high temperatures. However, a sharp decrease in buckling height is observed with temperature till 500 K due to the fact that the lattice constant is increasing (Fig. \ref{fig:effect_of_temp}(b)). The buckling height is apparently a constant in the range of 500 K - 750 K and slightly increases till 1000 K. Compared to the monolayer, the buckling height is less sensitive to temperatures beyond 500 K. We find that there is little change in the lattice constant with the temperature in bilayer compared to the monolayer. In Fig. \ref{fig:effect_of_temp}(c), we plot the \% deviation of the lattice constant with respect to the value at 0 K. The maximum observed deviation in the lattice constant is only 0.8\% at an elevated temperature of 1000 K. Counterintuitively, we observe that the lattice constant increases with temperature till 300 K and shows a decreasing trend till 500 K. This is not the case in the monolayer silicene, indicating interlayer interactions and changes in buckling height might collectively result in the unusual trends for lattice constant. In Fig. S2(a) (in the Supplemental Material \cite{supplement}), we find that as the temperature increases, there is a monotonic decrease in the lattice constant from $\sim$3.81 to $\sim$3.71 {\AA} as the temperature is increased from 0 K to 1000 K. However, it is well known that the lattice thermal expansion of 2D materials varies widely with the choice of interatomic potentials \cite{Herrero2020, Pozzoprl2011, Magnin2014, Lindsay2010}. An accurate description of these properties will require the use of \textit{ab-initio} molecular dynamics simulations (AIMD). We will address the behavior thermal expansion of bilayer silicene with accurate AIMD in future communication.  As expected, we observe significant corrugations in the monolayer and bilayer silicene (Fig. \ref{fig:crystal_structure}) at finite temperatures. The degree of corrugations is slightly suppressed in bilayer silicene compared to its monolayer counterpart. In the case of the bilayer, one layer of silicene acts as the substrate stabilizing the other layer, compatible with previous studies on silicene based heterostructures \cite{Geng2018}.

The electronic structure and phonon properties exhibit pronounced sensitivity to structural changes induced by temperature variations. To investigate the effect of temperature, we employ DFT to calculate the electronic band structure along the $\Gamma$$\rightarrow$K$\rightarrow$M$\rightarrow$$\Gamma$ direction (Fig. \ref{fig:effect_of_temp}(d)) for the structures predicted by classical simulations at 0 and 300 K. Surprisingly, our calculations indicate the presence of a Dirac cone at 0 K, contrary to previous findings \cite{Liu_2014}. This difference highlights a significant issue with the structures predicted by classical molecular mechanics. It serves to emphasize that the SW+LJ model, even after refinement in our study, is inadequate for predicting structures intended to be subsequently used when performing electronic structure calculations using DFT.

We also calculate the temperature effects on phonon dispersion in the same high symmetry direction as the electronic band structure in the $\bm{q}$-space (Fig. \ref{fig:effect_of_temp}(e)). We observe that both the optical and acoustic branch of the phonon spectrum undergoes red-shifting as the temperature is increased with a higher redshift in the optical phonon branch compared to the acoustic branch. More details about the phonon modes and the effect of temperature are discussed in Section \ref{subsec:phonon}. In summary, we see significant and unconventional changes in the structural, electronic, and phononic behavior of the bilayer compared to the monolayer counterpart with the increase in temperature. In the following section, we shed light on how the above results affect thermal transport in the bilayer system.

\subsection{\label{subsec:cond}Thermal Conductivity of Bilayer Silicene}
\begin{figure}[b]
	\includegraphics[height=6cm,keepaspectratio]{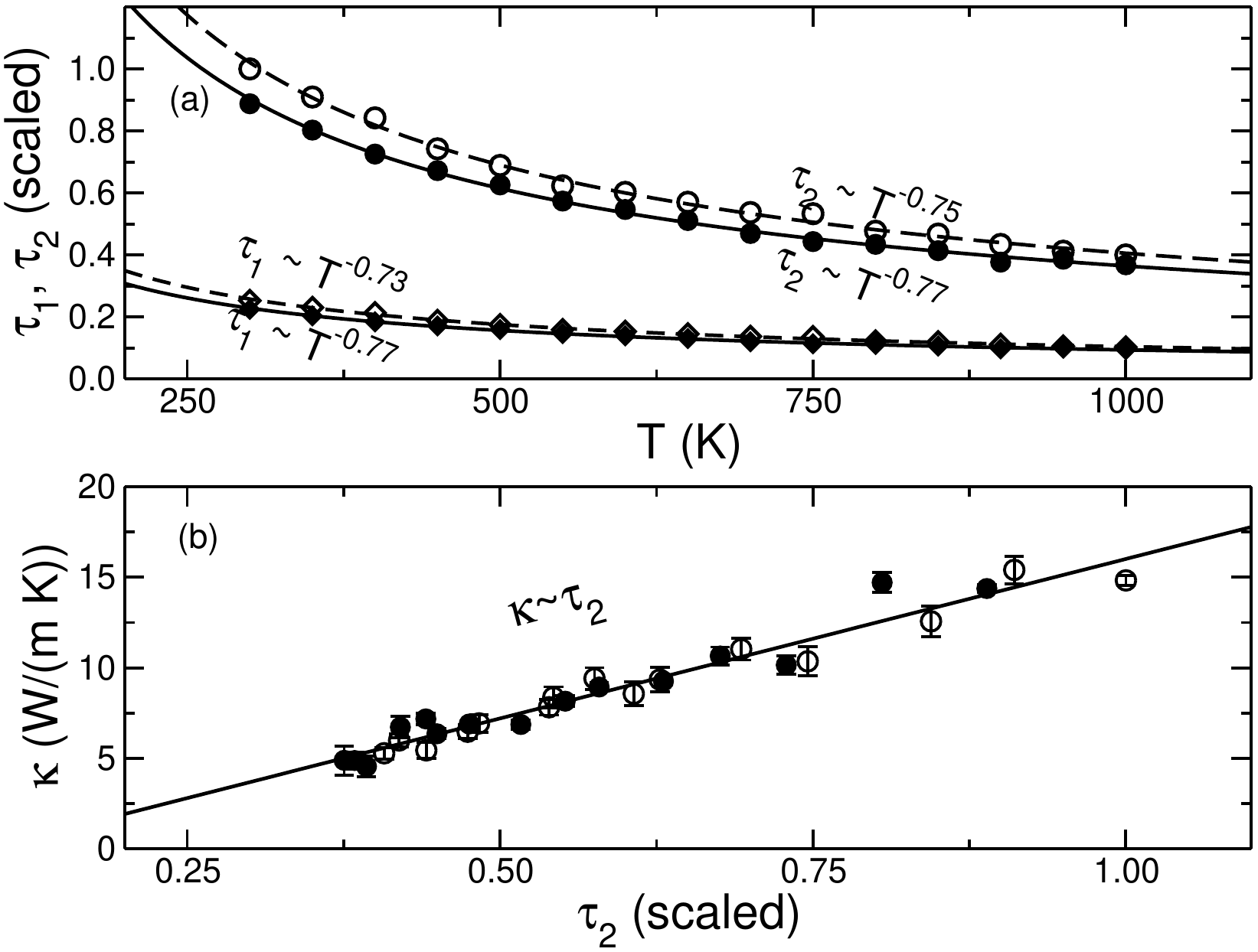}
	\caption{(a) The characteristic time scales associated with short-range ($\tau_1$) and long-range ($\tau_2$) correlations as a function of temperature. The filled-in symbols and solid lines represent the simulation data and the power-law fit for the monolayer system, respectively. The open symbols and the dashed lines represent the simulation data, and the power law fit for the bilayer system. Short-range characteristic time scales are shown in diamonds, and the long-range characteristic time scales are shown in circles. (b) Thermal conductivity as a function of the long-range characteristics time scales. Other legends are the same as that of Fig. \ref{fig:kappa_vs_T}. It is evident that the temperature has the same effect on $\tau$ as it has on $\kappa$, which explains convincingly that $\kappa\sim T^{-0.93}$.}
	\label{fig:char_time}
\end{figure}

\begin{figure}[b]
	\includegraphics[height=6cm,keepaspectratio]{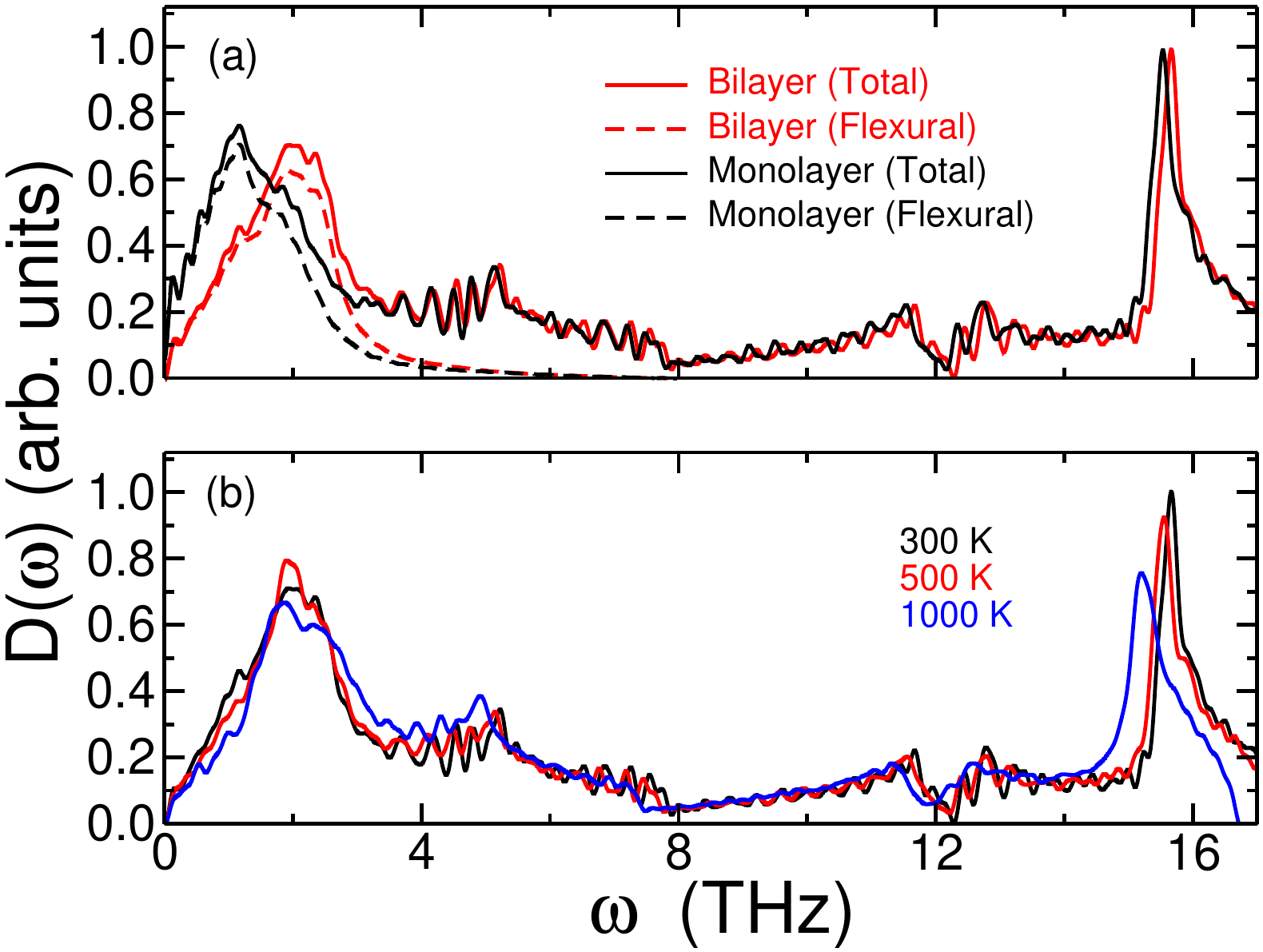}
	\caption{(a) Phonon DoS as a function of the angular frequency of vibration for monolayer (solid black curve) and bilayer (solid red curve) silicene. The absolute value is rescaled to better represent the dependency on the interactions. Flexural contribution is plotted with dashed lines (black and red for monolayer and bilayer, respectively). (b) Dependency of phonon DoS on temperature for bilayer silicene. Black, red, and blue curves represent 300 K, 500 K, and 1000 K, respectively. The intensity of phonon density of state decreases with temperature due to increased phonon scattering rate at high temperatures.}
	\label{fig:dos}
\end{figure}
The thermal transport properties of a material are characterized by thermal conductivity. Using the EMD simulations, we calculate the thermal conductivity of the silicene bilayer. The equilibrium approach uses the Greek-Kubo relation (Eq. \ref{eqn:gk_cond}), which integrates the heat current autocorrelation function (HCACF) to calculate the thermal conductivity. In Fig. \ref{fig:kappa_vs_T}(a), we calculate the normalized HCACF of bilayer silicene as a function of correlation time or lagtime, $\tau$. We observe that the correlation function decays exponentially with oscillations till $\sim$ 50 ps, beyond which there is a negligible contribution to thermal conductivity. For calculating the thermal conductivity, we simulated 30 independent trajectories with a length of 1 ns with a trajectory saving frequency of 0.02 fs. We note that the choice of the number of independent trajectories strictly depends on the convergence of the system and the desired accuracy. Fig. \ref{fig:kappa_vs_T}(b) shows the thermal conductivity calculated from individual runs and the average thermal conductivity based on 30 independent runs as a function of the upper limit of the GK integral. The thermal conductivity converges after about $\sim$20 ps of correlation time for 300 K. For higher temperatures, $\kappa$ converges faster, which can be attributed to the lower lifetime of phonons at higher scattering. However, to reduce errors, we have taken 100 ps as the cutoff correlation time, based on the decay of the heat current autocorrelation function for estimating the thermal conductivity.

Our calculations show that the thermal conductivity of the silicene bilayer is 14.8 $\pm$ 0.3 W/(m K) at room temperature, much lower than planer 2D materials such as graphene and h-BN. We see an unusual increase in the thermal conductivity of about $\sim 15\%$ in bilayer silicene compared to the monolayer system. This is contrary to previously reported trends for graphene and h-BN \cite{Kong2009prb, Ouyang2020acs}. A further understanding of this result requires a detailed analysis of phonons conducted in \ref{subsec:phonon}. We notice a monotonic decrease in the thermal conductivity of both the mono- and bilayer systems with temperature (Fig. \ref{fig:kappa_vs_T}(c)), similar to other 2D materials. We attribute this decrease in thermal conductivity to phonon-phonon Umklapp processes. At higher temperatures, the Umklapp process increases significantly, thus reducing the thermal conductivity. Since the thermal conductivity is ideally found to be inversely proportional to temperature, we fit $\kappa$ to a power-law and obtain $\kappa\sim T^{-0.93}$ for monolayer and $\kappa\sim T^{-0.88}$ for bilayer silicene, with an overall trend to be $\kappa\sim T^{-0.9}$. The deviation from the ideal scaling of $T^{-1}$ for both systems is only marginal. This deviation from ideal $T^{-1}$ law might be to the quasi-harmonic nature of the scattering resulting from higher order effects \cite{Tretiakov2004}. 

To further understand the deviation from ideal $\kappa\sim T^{-1}$ law, we fit the correlation functions with two exponential functions such that: 
\begin{equation}
	\frac{\langle \bm{S}(0).\bm{S}(t) \rangle}{\langle \bm{S}(0).\bm{S}(0) \rangle} = a_0\exp(-t/\tau_1) + (1-a_0)\exp(-t/\tau_2).
\end{equation}
The fitting gives two characteristic times, a short-range characteristic time, $\tau_1$, and a long-range characteristic time $\tau_2$). We find that $\tau_1$ is less than $\tau_2$ signifying a faster decay at a lower correlation time (below $\sim$5 ps). Further, both the $\tau_1$ and $\tau_2$ decrease with the increase in temperature (Fig. \ref{fig:char_time}(a)). This indicates a lower phonon mean free path with the increase in temperature \cite{Tretiakov2004}. The lowering of the phonon mean free path contributes to the decrease in thermal conductivity at higher temperatures. We analyze the correlation between characteristic times and temperature in Fig. \ref{fig:char_time}(a), and by fitting to a power-law, we see that for both the monolayer and bilayer systems scale as $\tau_1\sim T^{-0.75}$ and $\tau_2\sim T^{-0.77}$. To investigate the existence of any scaling relation between the characteristics times and the thermal conductivity, in Fig. \ref{fig:char_time}(b), we compare $\kappa$ and $\tau_2$, the long-range characteristic time as it contains more intricacies of phonons. It is found that the thermal conductivity has a linear relation with the characteristics time ($\kappa\sim\tau_2$) qualitatively. Although this simple analysis provides a significant outlook, to further understand the behavior of bilayer silicene compared to other 2D materials, we explore the dependency of phonon modes in systems with temperature in the subsequent section. We note that the quantum corrections are not incorporated in our calculations due to the unreliable estimate of the quantum effects using these techniques. No classical counterpart of zero point energy exists; hence every quantum correction is deemed unreliable \cite{Fan2017prb}. Since the main goal of this work is to compare and understand the effect of temperature and interactions on thermal transport, our classical results are valid for all considerations.


\subsection{\label{subsec:phonon}Phonons in Bilayer Silicene}
\begin{figure}[b]
	\includegraphics[height=6cm,keepaspectratio]{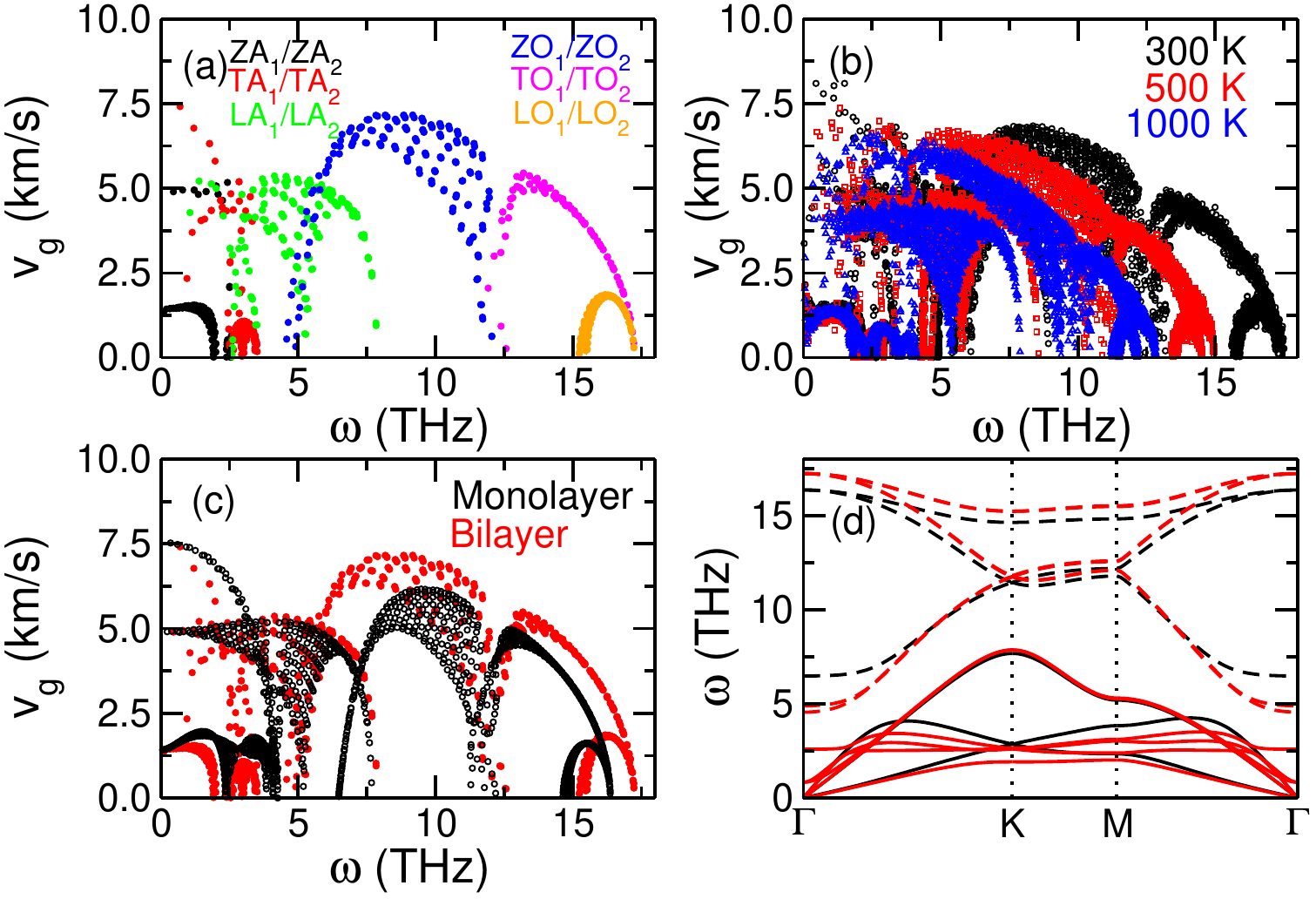}
	\caption{(a) Phonon group velocity with angular frequency in bilayer silicene at 300 K, (b) variation of the group velocity for the bilayer at 300 K, 500 K, 1000 K temperatures, (c) variation of the group velocity for monolayer (black circles) and bilayer (red circles) and (d) comparison of phonon band structure for monolayer (black curve) and bilayer (red curve) at 300 K. The acoustic branches are represented with solid curves while the optical branches are represented with dashed curves. We observe that group velocity decreases with increased temperature, leading to reduced thermal conductivity. However, LA$_1$/LA$_2$ mode velocities are increased in the bilayer compared to the monolayer leading to higher thermal conductivity.}
	\label{fig:fig7}
\end{figure}
To understand the effect of interlayer interactions and temperature on thermal conductivity, we have systematically analyzed the phonon dynamics in the silicene systems. The phonon density of states provides insights into how the phonon modes contribute to the thermal conductivity of the systems. We calculate the phonon density of states and the phonon partial density of states to understand how the phonon density is influenced by the in-plane and out-of-plane motion of the atoms. The phonon density of states, calculated from the MD simulations, is simply a Fourier transform of the VACFs. We plot the normalized VACF and the contribution of the out-of-plane component to the VACF as a function of the correlation or lagtime in Fig. S3(a) (in the Supplemental Material \cite{supplement}). We find that the VACF decays to zero after $\sim$ 10 ps of correlation time. Hence, the maximum time limit for the integration in Eq. (\ref{eqn:dw}) is taken as 50 ps, which is more than sufficient for all the fluctuations to die out. Interestingly, the out-of-plane component shows lesser fluctuations over the whole range of correlation time. We plot the rescaled phonon DoS for the silicene bilayer and monolayer system in Fig. \ref{fig:dos}(a). We note that the acoustic mode density blueshifts in the bilayer, which result in an increased thermal conductivity of the bilayer silicene compared to the monolayer. The flexural component of the DoS contributes to total DoS at lower frequencies but decays to zero at the optical frequencies. Nevertheless, both the acoustic and optical modes are blueshifted for the bilayer system. For understanding the effect of temperature on phonon DoS, the total phonon density of states for the bilayer is plotted at 300 K, 500 K, and 1000 K temperatures in Figure \ref{fig:dos}(b). It is explicit that there is a monotonic decrease in intensity in the phonon dispersion curve with temperature. The redshift in the peak on the phonon DoS (Fig. S3(b), in the Supplemental Material \cite{supplement}) directly correlates to the decrease in thermal conductivity with temperature for both the silicene systems \cite{santosh2018}. This softening of the phonon modes with temperatures is, in fact, due to the increase in the interatomic and interlayer separations.

In Fig. \ref{fig:fig7}(a), we have plotted the group velocity of the bilayer system at 300 K. From the calculations, it is found that the flexural ZA$_1$/ZA$_2$ phonon modes have less group velocity than the other acoustic modes, i.e., TA$_1$/TA$_2$ and LA$_1$/LA$_2$. This observation points to the fact that due to the inherent buckling, the flexural modes are prone to higher scattering than the in-plane modes and, thus, possess lower group velocity. Naturally, the thermal conductivity due to flexural modes is expected to be limited, unlike in the cases of graphene and h-BN \cite{Kong2009prb}. However, in silicene, the LA modes contribute the most to the thermal conductivity \cite{Wang2015}. Among the acoustic modes, the LA$_1$/LA$_2$ modes have higher group velocity in bilayer silicene (Fig. \ref{fig:fig7}(a)). We observe a blueshift of phonon branches apparent in the bilayer compared to the monolayer system (Fig. \ref{fig:fig7}(b)), similar to the blueshift seen in the phonon DoS. Overall, it is found that the bilayer phonon modes have higher group velocities compared to the monolayer system. Although bilayer silicene is found to possess lower flexural group velocity (Fig. \ref{fig:fig7}(b)), the much higher magnitude of the group velocity of the LA$_1$/LA$_2$ modes explains the unusually higher thermal conductivity of the bilayer silicene compared to the monolayer counterpart. With temperature, there is a monotonic decrease in the group velocity of the bilayer system, shown in Fig. \ref{fig:fig7}(c). As the temperature is raised, the scattering probability of phonons increases, leading to higher collisions which decrease the group velocity. The reduced group velocity leads to a lowered thermal conductivity of the systems, as discussed in the previous section. Clearly, the redshifting of phonon modes directly correlates with the decrease in thermal conductivity.

We use the classical molecular dynamics trajectories to calculate the phonon dispersion of the monolayer and bilayer systems. This approach of calculating the phonon dispersion agrees well with the phonon dispersion for the monolayer calculated using the density functional perturbation theory (DFPT) (Fig. S4(a) in the Supplemental Material \cite{supplement})) and the phonon spectral energy density (SED) method (Fig. S4(b) in the Supplemental Material \cite{supplement})) in the $\Gamma\rightarrow$ M direction. Since our method with the classical potentials reproduces the results of quantum calculations within a reasonable limit for uncertainties (Fig. S4(a) in the Supplemental Material \cite{supplement})), we use the classical mechanics based techniques for simplicity. The phonon dispersion curve calculated for the bilayer (Fig. S4(c) in the Supplemental Material \cite{supplement})) and the monolayer systems (Fig. S4(a) in the Supplemental Material \cite{supplement})) shows a similar trend. We notice a blueshift in both the optical and the acoustic branches of the bilayer compared to the monolayer, which directly corresponds to the increase in the thermal conductivity in the bilayer (Fig. \ref{fig:fig7}(d)). It is found that the in-plane and out-of-plane modes are strongly coupled in monolayer silicene due to inherent buckling \cite{swopt2014}. The flexural acoustic modes also experience a dampening effect due to the increased buckling induced out-of-plane scattering. However, because strong interlayer interactions exist between the silicene layers, the bonding leads to the generation of high-velocity LA$_1$/LA$_2$ elastic waves. Generally, this is observed in the case of substrate layer interactions, where increasing the interaction strength leads to higher thermal transport \cite{Ongpop2011, Guo2011}. This high velocity of the LA$_1$/LA$_2$ modes contributes the most to the thermal conductivity of silicene, thereby increasing the thermal conductivity of the bilayer compared to the monolayer system. To examine the effect of interaction strength on the group velocity of silicene bilayers, we have carried out a few simulation experiments by manually changing essentially the interaction parameter $\epsilon$ appearing in the LJ interaction potential. To implement this, we introduced a scaling parameter $\lambda$ in the LJ potential such that, $\phi\left(r_{i j}\right)=4 \lambda\epsilon\left[\left(\frac{\sigma}{r_{i j}}\right)^{12}-\left(\frac{\sigma}{r_{i j}}\right)^6\right]$. We have calculated the group velocities associated with the LA phonon modes of the bilayer silicene systems for two different values of $\lambda$ (i.e., 0.5 and 1.5), to change the strength of interaction by 50 \% weaker and 50 \% stronger with respect to the value of interaction parameter $\epsilon$ optimized in this work. In short, we observe that the group velocities of LA phonons have a positive correlation with the strength of interaction in the studied range of interaction strength (Fig. S5 in the Supplemental Material \cite{supplement})). Further, the thermal conductivity also exhibits similar changes with a reduced thermal conductivity of 12.9 $\pm$ 0.4 W/(m K) for $\lambda$ = 0.5, consistent with the changes observed for the group velocities. Thus, the enhancement of LA$_1$/LA$_2$ phonon group velocities induced by stronger interlayer interactions ultimately contributes to unusually higher thermal conductivity in the bilayer silicene.

\section{\label{sec:conclusions}Conclusions\protect}
In summary, we developed accurate parameters of pairwise LJ potential for modeling interlayer interactions between silicene layers and investigated the phonon thermal transport of mono- and bilayer silecene with the Green-Kubo approach. The developed set of parameters correctly predicts the interlayer separation between the layers and phonon band structure in agreement with the \textit{ab-initio} DFT calculations \cite{Liu2013}.

The interlayer interactions and temperature play a major role in determining the thermal transport properties of the silicene layered materials. Notably, the behavior of ZA$_1$/ZA$_2$ modes in bilayer silicene exhibits a reduction in group velocity when compared to the monolayer silicene systems. However, intriguingly, the thermal conductivity in bilayer silicene is found to be higher than that in monolayer silicene at any given temperature, which is attributed to the enhancement in group velocity of LA$_1$/LA$_2$ modes arising due to strong covalent interactions. Furthermore, the temperature-dependent variation of silicene's thermal conductivity follows a power-law relationship, consistent with the trends observed in other two-dimensional materials. This decrease in thermal conductivity follows a scaling trend of $\kappa\sim T^{-0.9}$, which diverges from the typically reported $\kappa\sim T^{-1}$ scaling, owing to the quasi-harmonic nature of phonon scattering at elevated temperatures.

Our investigation has revealed novel insights into the intricate phonon transport mechanisms inherent in bilayer silicene, facilitated by the development of interlayer pairwise interaction potential parameters. Specifically, our analysys establishes that the phonon thermal conductivity correlates excellently with the underlying relaxation timescales associated with the decay of heat current autocorrelation functions. In addition to understanding the phonon mechanisms in slicene systems, our study may pave forward their potential utility in thermal interfacing devices and thermoelectric applications. We propose further experimental as well as theoretical studies at first-principles level to further establish the complete nature of thermal transport in silicene.

\begin{acknowledgments}
The authors acknowledge the Computer Center of IIT Jodhpur and the HPC Center at the Department of Physics, Freie Universität Berlin (10.17169/refubium-26754), for providing computing resources that have contributed to the research results reported in this paper. SM acknowledges support from the SERB International Research Experience Fellowship SIR/2022/000786 and SERB CRG/2019/000106 provided by the Science and Engineering Research Board, Department of Science and Technology, India. SSPC acknowledges the Ministry of Education (MoE), Govt. of India, for the financial support received as a fellowship. AS acknowledges support from the SERB through grant number RJF/2021/000147.
\end{acknowledgments}


\bibliography{bl_silicene_manuscript}

\end{document}